\documentclass[twocolumn,showpacs,amsmath,amssymb]{revtex4}
\input{epsf}

\usepackage{graphicx}
\usepackage{array}
\usepackage{bigstrut}
\usepackage{longtable}
\usepackage{rotating,booktabs}
\usepackage{booktabs,threeparttable}
\usepackage{bm}
\usepackage{lipsum}
\usepackage{warpcol}
\usepackage{color}

\begin{document}

\title{ Proposal for suppressing ac Stark shift in the He($2\,^3S_1\rightarrow 3\,^3S_1$) two-photon transition using magic wavelengths}
%\title{ Possibility of improving the $2\,^3S_1\rightarrow 3\,^3S_1$ transition spectroscopy of helium by using magic wavelengths }

\author{Yong-Hui Zhang, Li-Yan Tang, and Ting-Yun Shi} %$^{\dag}$~\footnotetext{$\dag$Email Address: tyshi@wipm.ac.cn}

\affiliation {State Key Laboratory of Magnetic Resonance and Atomic and Molecular Physics, Wuhan Institute of Physics and
Mathematics, Innovation Academy for Precision Measurement Science and Technology, Chinese Academy of Sciences, Wuhan 430071, People's Republic of China}

%\affiliation {$^2$ Center for Cold Atom Physics, Chinese Academy of Sciences,
%Wuhan 430071, People¡¯s Republic of China}

\date{\today}

\begin{abstract}

Motivated by recent direct measurement of the forbidden $2\,^3S_1\rightarrow3\,^3S_1$ transition in helium [K. F. Thomas et al., Phys. Rev. Lett. 125, 013002(2020)], where the ac Stark shift is one of the main systematic uncertainties, we propose a dichroic two-photon transition measurement for $2\,^3S_1\rightarrow3\,^3S_1$ which could effectively suppress the ac Stark shift by utilizing magic wavelengths: one magic wavelength is used to realize state-insensitive optical trapping, the other magic wavelength is used as one of the two lasers driving the two-photon transition.
Carrying out calculations based on the no-pair Dirac-Coulomb-Breit Hamiltonian with mass shift operator included, we report the magic wavelength of 1265.615 9(4) nm for $^4$He [or 1265.683 9(2) nm for $^3$He] can be used to design an optical dipole trap; the magic wavelength of 934.234 5(2) nm for $^4$He [or 934.255 4(4) nm for $^3$He] can be as one excitation laser in the two-photon process, and the ac Stark shift can be reduced to less than 100 kHz, as long as the intensity of the other excitation laser does not exceed $1\times10^4~W/cm^2$.
Alternatively, by selecting detuning frequencies relative to the $2\,^3P$ state in the region of 82$\sim$103 THz, as well as adjusting the intensity ratios of the two lasers, the ac Stark shift in the $2\,^3S_1\rightarrow3\,^3S_1$ two-photon transition can be cancelled.

\end{abstract}

  \maketitle

\section{introduction}

High-precision absolute frequency measurements in helium provide an ideal platform for testing QED theory and determining fundamental constants, such as the fine structure constant and nuclear charge radius~\cite{pachucki17a, patkos20, patkos21,Zheng17a, kato18, pastor12a, zheng17, wim18a}, which benefits from the abundant laser accessible transition spectra of helium itself. Table~\ref{t0} summarizes four transitions from the long-lived metastable $2\,^3S_1$ state to other excited states in $^4$He. It is seen that the most precise frequency measurement of helium has reached ppt ($10^{-12}$) level~\cite{wim18a} for the $2\,^3S_1\rightarrow 2\,^1S_0$ transition. Compared with other three transitions, the uncertainty of 5 MHz~\cite{thomas20} for the transition frequency of an ultraweak $2\,^3S_1\rightarrow 3\,^3S_1$ transition in $^4$He, which has been measured by single-photon transition, could be further reduced. One of the main systematic uncertainties is due to the ac Stark shift caused by the probe beam with an laser intensity at the focus of 3.86$\times 10^3~W/cm^2$, that is 6.9 MHz (exceeds the natural linewidth of 4.43 MHz)~\cite{wiese09}. In order to improve the measured precision for the $2\,^3S_1\rightarrow 3\,^3S_1$ transition, suppressing the ac Stark shift effectively becomes a major task in the future experiment.
\begingroup
\squeezetable
\begin{table}[!htbp]%\scriptsize
\caption{\label{t0} The lifetime of the upper state $\tau_{u}$, the theoretical natural linewidth $\Gamma$, the transition type, the transition rate $A$, and the current experimental measurement precision for four transitions from the $2\,^3S_1$ state of $^4$He. }
\begin{ruledtabular}
 \begin{tabular}{lrrccl}
 \multicolumn{1}{c}{Transitions}    &\multicolumn{1}{c}{$\tau_{u}$~\cite{wiese09}}     &\multicolumn{1}{c}{$\Gamma$}
&\multicolumn{1}{c}{Type} &\multicolumn{1}{c}{$A(s^{-1})$} &\multicolumn{1}{l}{Uncertainty}\\
\hline
$2\,^3S_1\rightarrow 2\,^3P$&97.9 ns  &1.63 MHz &E1&$10^7$          &1.4 kHz~\cite{zheng17} \\
$2\,^3S_1\rightarrow 2\,^1S_0$&20 ms    &7.96 Hz  &M1&$6.1\times10^{-8}$&0.2 kHz~\cite{wim18a}      \\
$2\,^3S_1\rightarrow 2\,^1P_1$&0.56 ns  &287 MHz  &E1&1.442&0.5 MHz~\cite{notermans14a}\\
$2\,^3S_1\rightarrow 3\,^3S_1$&35.9 ns  &4.43 MHz &M1&$6.48\times10^{-9}$&5 MHz~\cite{thomas20}        \\
\end{tabular}
\end{ruledtabular}
\end{table}
\endgroup

Reducing the ac Stark shift in a single-photon process for the $2\,^3S_1\rightarrow 3\,^3S_1$ transition might be challenging. The probe laser wavelength 427.7 nm~\cite{thomas20} has exceeded the wavelength 663.4 nm of the $3\,^3S_1$ state ionization energy, which will cause the ionization and decrease the population distribution of the $3\,^3S_1$ state. Correspondingly, the detection efficiency of this transition would be affected. We can reduce the frequencies of probe beams through a two-photon process to avoid the ionization, and the ac Stark shift in two-photon transitions can be suppressed by utilizing two lasers with different wavelengths $\lambda_1$ and $\lambda_2$, which has been described and realized in rubidium~\cite{perrella13, martin19, gerginov18, perrella19}, antiproton helium~\cite{hori10, hori11}, and molecular hydrogen ion~\cite{tran03, patra20}. In the present work of helium, we further propose one of the two lasers ($\lambda_2$ laser) is set to be a magic wavelength, at which the $2\,^3S_1$ and $3\,^3S_1$ states have the same dynamic polarizability, then we can minimize the total ac Stark shift in the $2\,^3S_1\rightarrow 3\,^3S_1$ two-photon transition only by carefully controlling the intensity of the $\lambda_1$ laser.

Furthermore, the single-photon process for the $2\,^3S_1\rightarrow 3\,^3S_1$ transition, which is excited via the magnetic dipole (M1) interaction, is an ultraweak transition since the Einstein $A$ coefficient is at the level of 10$^{-9}$ to 10$^{-8}$ $s^{-1}$~\cite{thomas20, derevianko98, lach01}. The ultraweak transitions can be detected in an optical dipole trap (ODT), moreover systematic uncertainties can be simultaneously reduced and characterized to kHz level~\cite{rooij11a, notermans14a}. For example, the most accurate measurement of the $^4$He double forbidden $2\,^3S_1\rightarrow 2\,^1S_0$ transition so far benefits from the use of a magic wavelength ODT~\cite{notermans14b, wim18a}. Based on this, we would also expect to probe the $2\,^3S_1\rightarrow 3\,^3S_1$ two-photon transition in an ODT, preferably operated at a magic wavelength. On the one hand, the lowest order ac Stark shift of the trapping laser can be cancelled; on the other hand, it is helpful to reduce the Doppler shift in the two-photon process.

In this work, for the $2\,^3S_1\rightarrow 3\,^3S_1$ transition spectral measurement, we propose to use a magic wavelength ODT to trap helium atoms, and use two different-wavelength lasers to realize the two-photon excitation with one of them is set to be a magic wavelength. To validate the feasibility of the present scheme, the required laser power for trapping helium atoms and the scattering rate limiting the transition coherence lifetime are evaluated; the ac Stark shift is analysed by controlling the intensity of $\lambda_1$ when $\lambda_2$ is a magic wavelength. We find that the ac Stark shift in the $2\,^3S_1\rightarrow 3\,^3S_1$ two-photon transition can be suppressed to less than 100 kHz, which paves a new way for improving the measured precision of the $2\,^3S_1\rightarrow 3\,^3S_1$ transition frequency. We also find that with appropriate detuning frequencies, the ac Stark shift in the $2\,^3S_1\rightarrow 3\,^3S_1$ two-photon transition can be minimized to zero by adjusting the laser intensity ratios. Atomic units (a.u.) are used throughout this paper unless stated otherwise.

\section{details of the calculations}

The relativistic energies and wave functions of helium are obtained using B-splines relativistic configuration interaction (RCI) method that has been described in our previous papers~\cite{zhang19, wu18, zhang21}. The RCI calculations are carried out by solving the eigenvalue problem of the no-pair Dirac-Coulomb-Breit (DCB) Hamiltonian with mass shift (MS) operator included. The two-electron configuration-state functions are constructed by the positive-energy single-electron wave functions with the orbital angular momentums less than the maximum partial wave $\ell_{max}$. The single-electron wave functions are acquired by solving the Dirac equation using Notre Dame basis sets of $N$ B-spline functions of order $k=7$~\cite{johnson88a, tang12b}. The nuclear mass $m_0$ of $^4$He and $^3$He are respectively $m_0=7294.299\,5361\, m_e$ and $m_0=5495.885\,2754\, m_e$~\cite{mohr12a}, where $m_e$ is the electron mass.

%=======================================================================================================
%c.\textit{Dynamic polarizability formula.}
%=======================================================================================================
%To the leading order in laser intensity and the fine-structure constant,
Magic wavelengths are located by calculating the dynamic dipole polarizabilities of two states involved in the atomic transition and finding their crossing points. The dynamic dipole polarizability of the magnetic sublevel $|N_gJ_gM_g\rangle$ under the linear polarized light with laser frequency $\omega$ is given by~\cite{mitroy10a, zhang16a, wu18}
\begin{eqnarray}
\alpha_1(\omega)=\alpha_1^S(\omega)+\dfrac{3M_g^2-J_g(J_g+1)}{J_g(2J_g-1)}\alpha_1^T(\omega)
\,,\label{e3}
\end{eqnarray}
where the scalar dipole polarizabilities $\alpha_1^S(\omega)$ is written as
\begin{eqnarray}
\alpha_1^S(\omega)=\sum_{n\neq g}\dfrac{2|\langle N_gJ_g\|T^{(1)}
                                \|N_nJ_n\rangle|^2\Delta E_{gn}}{3(2J_g+1)\left(\Delta E_{gn}^2-\omega^2\right)}
\,,\label{e4}
\end{eqnarray}
and the tensor dipole polarizabilities $\alpha_1^T(\omega)$ is defined as
\begin{widetext}
\begin{eqnarray}
&\alpha_1^T(\omega)=\sum\limits_{n\neq g}
(-1)^{J_g+J_n}\sqrt{\dfrac{30(2J_g+1)J_g(2J_g-1)}{(2J_g+3)(J_g+1)}} %\nonumber \\
&\left\{
\begin{array}{ccc}
1   &1   &2\\
J_g &J_g &J_n\\
\end{array}
\right\} \dfrac{2|\langle N_gJ_g\|T^{(1)}
                                \|N_nJ_n\rangle|^2\Delta E_{gn}}{3(2J_g+1)\left(\Delta E_{gn}^2-\omega^2\right)}
\,,\label{e5}
\end{eqnarray}
\end{widetext}
with $\triangle E_{gn}=E_n-E_g$ being transition energy between the initial state $|N_gJ_g\rangle$ and the intermediate state $|N_nJ_n\rangle$, and $T^{(1)}=\sum_{i=1}^2r_iC^{(1)}(\hat{r}_i)$ being the electric dipole transition operator in the length gauge.

%========================================================================================================
%a.\textit{trap parameter}
The potential depth $U$ of an ODT and the scattering rate $\Gamma_{sc}$ of atoms can be written in terms of the dynamic dipole polarizabiliy $\alpha_1(\omega)$~\cite{grimm00, notermans14b, porsev04a},
\begin{eqnarray}
U=\dfrac{1}{\varepsilon_0 c_0}\alpha_1(\omega)\dfrac{2P}{\pi w_0^2}\,,\label{e6}
\end{eqnarray}
\begin{eqnarray}
\Gamma_{sc}=\dfrac{4}{3h\varepsilon_0^2 c_0^4}\omega^3\alpha_1^2(\omega)I_0\,,\label{e7}
\end{eqnarray}
where $\varepsilon_0$, $c_0$, and $h$ are the dielectric constant, the speed of light in vacuum, and Planck constant, respectively, $P$ is the power of the trapping laser beam, $w_0$ is the beam waist, $I_0=2P/(\pi w_0^2)$ is the laser intensity, and $\omega$ is the angular frequency of the trapping light. For a magic wavelength ODT, $\omega$ is the magic frequency. The physical quantities in Eqs.~(\ref{e6}) and (\ref{e7}) are in the International System of Units (SI).

%========================================================================================================
%a.\textit{two photon}
The two-photon electric dipole (2E1) differential decay rate of the upper state $|e\rangle$ (or $|N_eJ_eM_e\rangle$) to the lower state $|g\rangle$ is given by~\cite{safronova10e, derevianko97, bondy20}
\begin{eqnarray}
\dfrac{dA^{2E1}}{d\omega_1}=\dfrac{8}{9\pi}\alpha^6\omega_1^3\omega_2^3\sum\limits_{q_1q_2}|M^{2E1}_{q_1q_2}|^2\,,\label{e11}
\end{eqnarray}
where $\alpha=1/137.035999074$~\cite{mohr12a} is the fine structure constant, the photon frequencies obey energy conservation, $\omega_1+\omega_2=E_e-E_g$, and the two-photon transition matrix element $M^{2E1}_{q_1q_2}$ can be expressed as
\begin{widetext}
\begin{eqnarray}
M^{2E1}_{q_1q_2}=\sum\limits_n\left[\dfrac{\langle e|T^{(1)}_{q_2}|n\rangle\langle n|T^{(1)}_{q_1}|g\rangle}{E_n+\omega_2-E_e}
+\dfrac{\langle e|T^{(1)}_{q_1}|n\rangle\langle n|T^{(1)}_{q_2}|g\rangle}{E_n+\omega_1-E_e}\right]\,\label{e12},%\nonumber \\
\end{eqnarray}
\end{widetext}
with $|n\rangle$ designating intermediate states and $T^{(1)}_{q_i}\,(i=1,2)$ being the $q_i$-th component of the electric dipole transition operator. Using Wigner-Eckart theorem, we perform summations over $q_1$, $q_2$ and magnetic quantum numbers of $M_g$ and $M_e$, then we obtain the following expression for the square of transition amplitude $|M^{2E1}|^2$,
%$M^2=\sum\limits_{q_1q_2}|M^{2E1}_{q_1q_2}|^2$ can be written as
%
\begin{widetext}
\begin{eqnarray}
|M^{2E1}|^2%=\sum\limits_{M_{g}M_e}\sum\limits_{q_1q_2}|M^{2E1}_{q_1q_2}|^2 %\nonumber \\
   =\sum\limits_{nn'J}\dfrac{1}{2J+1}\Big[T_{nJ}(\omega_2)T_{n'J}(\omega_2)+T_{nJ}(\omega_1)T_{n'J}(\omega_1)\Big]
  +2\sum\limits_{nJ_n}\sum\limits_{n'J_{n'}}
  \left\{
  \begin{array}{ccc}
  J_{n'} &1 &J_e\\
  J_n  &1 &J_g\\
  \end{array}
  \right\}
  (-1)^{J_n+J_{n'}}T_{nJ_n}(\omega_2)T_{n'J_{n'}}(\omega_1) \,,\nonumber \\\label{e13}
\end{eqnarray}
\end{widetext}
where
\begin{eqnarray}
T_{nJ_n}(\omega)=\dfrac{\langle N_eJ_e\|T^{(1)}\|N_nJ_n\rangle\langle N_nJ_n\|T^{(1)}\|N_gJ_g\rangle}{E_{n}+\omega-E_e}\,.\label{e14}
\end{eqnarray}

\section{results and discussions}
\subsection{dipole polarizabilities}

We use a complete set of configuration wave functions on an exponential grid~\cite{bachau01a} generated using B-splines constrained to a spherical cavity. A cavity radius of 200 a.u. is chosen to accommodate the initial state and the corresponding intermediate states, and it is also suitable for obtaining dynamic dipole polarizabilities of the $2\,^3S_1$ and $3\,^3S_1$ states for $\omega<0.068$ a.u., which corresponds to the ionization energy of the He($3\,^3S_1$) state. The basis set consists of $N=$ 40, 45, and 50 splines for each value of the partial wave that is less than $\ell_{max}$=10. The numerical uncertainty is evaluated by doubling the maximum difference between the extrapolated value and those given in the last three larger basis sets of convergence Tables.

\begin{table*}[!htbp]
\caption{\label{t1} Convergence of the scalar and tensor components (in a.u.), $\alpha_1^S(0)$ and $\alpha_1^T(0)$, the total static dipole polarizabilities (in a.u.) of $\alpha_1(0)(M=0)$ and $\alpha_1(0)(M=\pm1)$ for the $3\,^3S_1$ state in $^4$He and $^3$He from RCI calcualtions. The numbers in parentheses are numerical convergence uncertainties. }
\begin{ruledtabular}
\begin{tabular}{cllll}
 \multicolumn{1}{c}{($\ell_{max}$, N)}    &\multicolumn{1}{c}{$\alpha_1^S(0)$}     &\multicolumn{1}{c}{$\alpha_1^T(0)$}
&\multicolumn{1}{c}{$\alpha_1(0)(M=0)$}      &\multicolumn{1}{c}{$\alpha_1(0)(M=\pm1)$}\\
\hline
\multicolumn{5}{c}{$^4$He} \\
(9, 40)   &7940.358 781  &0.097 359  &7940.164 062 &7940.456 142\\
(10, 40)  &7940.361 519  &0.097 352  &7940.166 815 &7940.458 872\\
(10, 50)  &7940.359 702  &0.097 547  &7940.164 608 &7940.457 249\\
Extrap.   &7940.361(4)   &0.097(2)   &7940.166(4)  &7940.458(4) \\
\multicolumn{5}{c}{} \\
\multicolumn{5}{c}{$^3$He} \\
(9, 40)   &7942.117 552  &0.097 502  &7941.922 548 &7942.215 054\\
(10, 40)  &7942.120 509  &0.097 304  &7941.925 900 &7942.217 814\\
(10, 50)  &7942.120 877  &0.097 174  &7941.926 530 &7942.218 051\\
Extrap.   &7942.121(6)   &0.097(2)   &7941.927(8)  &7942.218(6)\\
\end{tabular}
\end{ruledtabular}
\end{table*}

As the numbers of the partial wave and B-splines increased, the convergence studies for the scalar and tensor polarizabilities, $\alpha_1^S(0)$ and $\alpha_1^T(0)$, and the static dipole polarizabilities of $\alpha_1(0)(M=0)$ and $\alpha_1(0)(M=\pm1)$ for the $3\,^3S_1$ state in $^4$He and $^3$He are presented in Table~\ref{t1}. For $^4$He, present RCI value of $\alpha_1^S(0)$ is 7940.361(4) a.u. with six convergent figures. Compared with the static polarizability of 7937.58(1) a.u.~\cite{yan00b} for $^\infty$He, it indicates that the static dipole polarizability is increased by 2.78 a.u. due to the finite nuclear mass and relativistic effects. These effects on $^3$He are more obvious than $^4$He.

\subsection{Determination of magic wavelengths}

\begin{figure}[!htbp]
\includegraphics[width=0.49\textwidth]{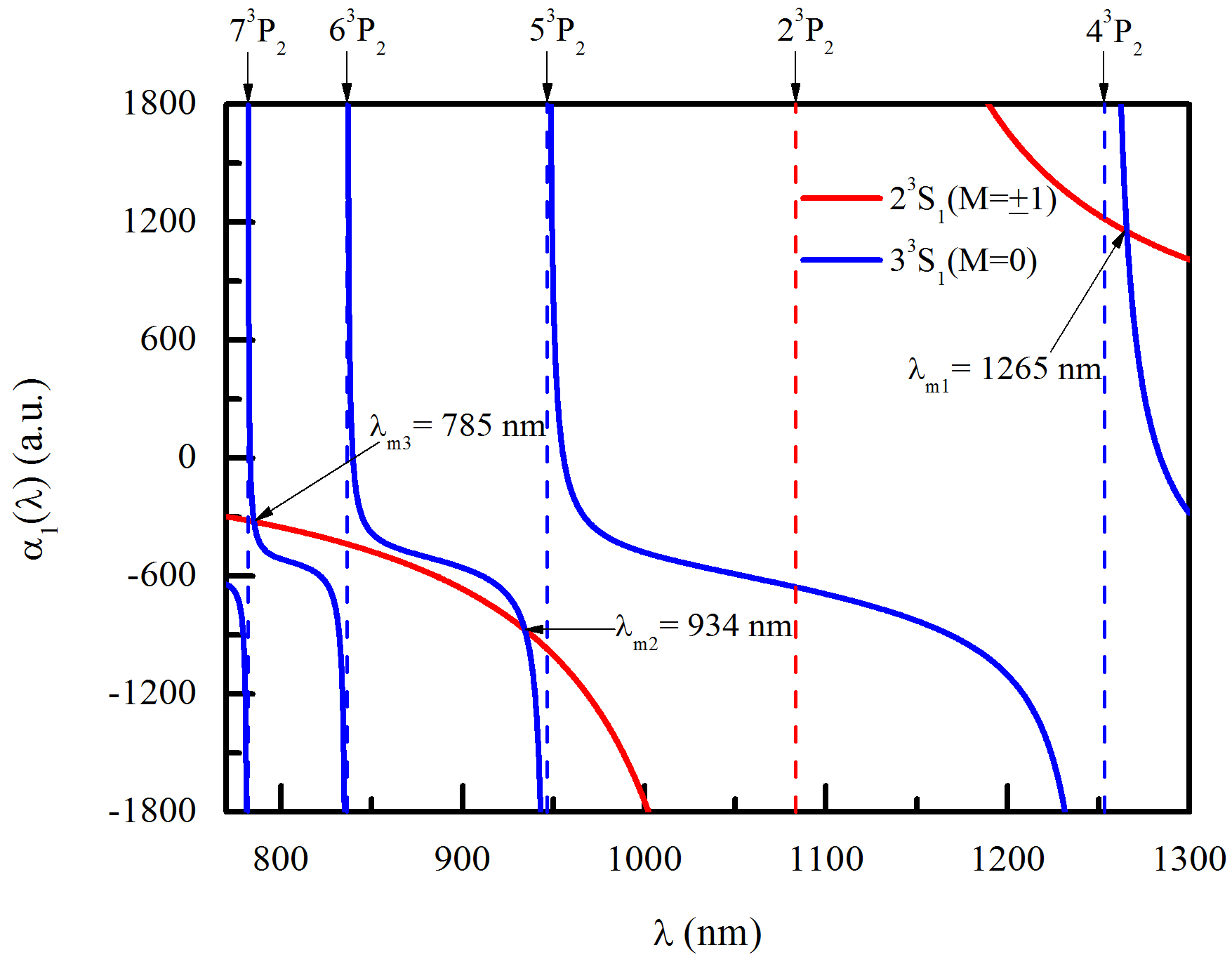}
\caption{\label{f1}(Color online) The dynamic dipole polarizabilities for the $2\,^3S_1(M=\pm1)$ and $3\,^3S_1(M=0)$ states of $^4$He. The magic wavelengths are marked with arrows. The positions of the resonances are indicated by vertical dashed lines with small arrows on top of the graph. }
\end{figure}

The magic wavelengths for the $2\,^3S_1\rightarrow 3\,^3S_1$ transitions need to be determined separately for the $M=0$ and $M=\pm1$ cases, since the total dynamic dipole polarizabilities for the $2\,^3S_1$ and $3\,^3S_1$ states depend upon the magnetic quantum numbers $M$. Fig.~\ref{f1} is the dynamic dipole polarizabilities in the range of 770$-$1300 nm for the $2\,^3S_1(M=\pm1)$ and $3\,^3S_1(M=0)$ magnetic sublevels of $^4$He. There are five resonances ($2\,^3S_1\rightarrow2\,^3P_2$, $3\,^3S_1\rightarrow4\,^3P_2$, $3\,^3S_1\rightarrow5\,^3P_2$, $3\,^3S_1\rightarrow6\,^3P_2$, and $3\,^3S_1\rightarrow7\,^3P_2$) existing in this wavelength region, which are indicated by vertical dashed lines with small arrows on top of the graph. Three magic wavelengths of $\lambda_{m1}$, $\lambda_{m2}$, and $\lambda_{m3}$ around 1265, 934, and 785 nm for the $2\,^3S_1(M=\pm1)\rightarrow 3\,^3S_1(M=0)$ transition are all marked with arrows in Fig.~\ref{f1}.

\begin{table*}[!htbp]
\caption{\label{t3} Magic wavelengths (in nm) and the corresponding dynamic dipole polarizabilities (in a.u.) obtained from RCI calculations for the $2\,^3S_1\rightarrow 3\,^3S_1$ transition of $^4$He and $^3$He, and the numbers in parentheses are numerical convergence uncertainties. For each transition, the first line refers to $^4$He, and the second line to $^3$He. }%For the $\alpha^3$-order QED correction, the numbers in parentheses are the final computational uncertainties to account for contributions from the electric-field dependence of the Bethe logarithm. For the $\alpha^4$-order QED correction, the figures listed are all convergent.}
\begin{ruledtabular}
\begin{tabular}{lllllll}
 \multicolumn{1}{c}{Transition}
&\multicolumn{1}{c}{$\lambda_{m1}$}  &\multicolumn{1}{c}{$\alpha_1(\lambda_{m1})$}
&\multicolumn{1}{c}{$\lambda_{m2}$}  &\multicolumn{1}{c}{$\alpha_1(\lambda_{m2})$}
&\multicolumn{1}{c}{$\lambda_{m3}$}  &\multicolumn{1}{c}{$\alpha_1(\lambda_{m3})$}\\
 \hline
$2\,^3S_1(M=0)\rightarrow3\,^3S_1(M=0)$       &1265.617 5(2)&1151.339(4)  &934.243 5(2) &$-$872.234(6)&785.327 4(2) &$-$324.405(2) \\
                                              &1265.691 6(2)&1151.954(6)  &934.270 1(6) &$-$871.75(2) &785.366(2)   &$-$324.363(2)\\
$2\,^3S_1(M=0)\rightarrow3\,^3S_1(M=\pm1)$    &1265.625 8(2)&1151.298(4)  &934.245 0(2) &$-$872.245(6)&785.329 5(2) &$-$324.409(2)\\
                                              &1265.699 9(2)&1151.913(6)  &934.271 6(6) &$-$871.76(2) &785.369(2)   &$-$324.366(2)\\
$2\,^3S_1(M=\pm1)\rightarrow3\,^3S_1(M=0)$    &1265.615 9(4)&1151.590(4)  &934.234 5(2) &$-$871.951(4)&785.326 8(2) &$-$324.362(2)\\
                                              &1265.689 9(2)&1152.205(6)  &934.261 2(4) &$-$871.47(2) &785.366(2)   &$-$324.320(2)\\
$2\,^3S_1(M=\pm1)\rightarrow3\,^3S_1(M=\pm1)$ &1265.624 2(2)&1151.549(4)  &934.236 0(2) &$-$871.962(4)&785.328 8(2) &$-$324.366(2)  \\
                                              &1265.698 2(2)&1152.164(6)  &934.262 7(4) &$-$871.48(2) &785.368(2)   &$-$324.323(2)\\
% \multicolumn{7}{c}{}\\
%$\alpha^3$-order QED correction               &0.006 34(6)  &0.109(1)   &$-$0.001 03(1)&0.117(1)  &0.002 40(2)&0.015 6(2)\\
%$\alpha^4$-order QED correction               &0.000 110    &0.001 9    &$-$0.000 018  &0.002 0   &0.000 042  &0.000 27\\
\end{tabular}
\end{ruledtabular}
\end{table*}

In previous work~\cite{zhang15}, the magic wavelengths for the $2\,^3S\rightarrow 3\,^3S$ transition of $^{\infty}$He have been determined based on NRCI calculations. In the present RCI calculations, we take account of the finite nuclear mass and relativistic effects on magic wavelengths and dynamic dipole polarizabilities. The updated results of $^4$He and $^3$He are shown in Table~\ref{t3}.

%\section{hyperfine interactions on $^3$He}
%
\begin{table*}[!htbp]
\caption{\label{t4} Magic wavelengths (in nm) and the corresponding dynamic polarizabilities (in a.u.) for the hyperfine transition of $2\,^3S_1, F=3/2, \rightarrow 3\,^3S_1, F=3/2$ of $^3$He. The numbers in parentheses are numerical convergence uncertainties.}
\begin{ruledtabular}
\begin{tabular}{cllllll}
 \multicolumn{1}{c}{Transition}
&\multicolumn{1}{c}{$\lambda_{m1}$} &\multicolumn{1}{c}{$\alpha_1(\lambda_{m1})$}
&\multicolumn{1}{c}{$\lambda_{m2}$} &\multicolumn{1}{c}{$\alpha_1(\lambda_{m2})$}
&\multicolumn{1}{c}{$\lambda_{m3}$} &\multicolumn{1}{c}{$\alpha_1(\lambda_{m3})$}\\
 \hline
$2\,^3S_1(M_F=\pm1/2)\rightarrow 3\,^3S_1(M_F=\pm1/2)$ &1265.685 1(2)  &1152.037(6)     &934.261 9(4)  &$-$871.65(2)     &785.363(2)    &$-$324.347(2)\\
$2\,^3S_1(M_F=\pm1/2)\rightarrow 3\,^3S_1(M_F=\pm3/2)$ &1265.682 0(2)  &1152.053(8)     &934.256 8(6)  &$-$871.61(2)     &785.361(2)    &$-$324.343(2)\\
$2\,^3S_1(M_F=\pm3/2)\rightarrow 3\,^3S_1(M_F=\pm1/2)$ &1265.683 9(2)   &1152.221(6)    &934.255 4(4)  &$-$871.44(2)      &785.362(2)     &$-$324.316(2)\\
$2\,^3S_1(M_F=\pm3/2)\rightarrow 3\,^3S_1(M_F=\pm3/2)$ &1265.680 7(2)   &1152.236(6)    &934.250 3(6)  &$-$871.40(2)      &785.361(2)     &$-$324.312(2)\\
\end{tabular}
\end{ruledtabular}
\end{table*}

Furthermore, for $^3$He, we consider the hyperfine interactions on magic wavelengths and polarizabilities. The dipole matrix elements between different hyperfine levels are transformed from the present matrix elements using the Wigner-Eckart theorem~\cite{jiang13c}. The hyperfine energy shifts given in Table 7 of Ref.~\cite{morton06b} are added to the present RCI energies of $2\,^3S_1$, $3\,^3S_1$, and $n\,^{1,3}P_{1,2,3}\, (n\leq10)$ to obtain the corresponding hyperfine energies. The dynamic polarizabilities for different hyperfine magnetic sublevels $|FM_F\rangle$ are calculated using Eqs.~(\ref{e3})-(\ref{e5}) by replacing $J$ and $M$ with $F$ and $M_F$, respectively. Then polarizabilities and magic wavelengths for different hyperfine transitions can be determined. Table~\ref{t4} shows magic wavelengths and the corresponding dynamic polarizabilities for different hyperfine transitions of $^3$He. Compared with the RCI calculated results of Table~\ref{t3}, it is seen that magic wavelengths become shorter due to the hyperfine interactions, and the dynamic dipole polarizabilities at the corresponding magic wavelengths are increased.

\subsection{1265 nm magic wavelength to design an ODT}

\begin{table}[!htbp]
\caption{\label{t5} The 1265 nm magic wavelengths of $^4$He ($2\,^3S_1, M=\pm1\rightarrow3\,^3S_1, M=0$) and $^3$He ($2\,^3S_1, F=3/2, M_F=\pm3/2\rightarrow 3\,^3S_1, F=3/2, M_F=\pm1/2$), dynamic dipole polarizabilities, and the photon recoil energies for $^4$He and $^3$He. Supposing that a 20$E_{r}$ trapping depth is created and the focused laser has a beam waist of 100 $\mu$m, the required laser beam power P and the corresponding trapping lifetime $\tau_{sc}=1/\Gamma_{sc}$ are given in the last two lines. The numbers in parentheses are numerical convergence uncertainties.}
\begin{ruledtabular}
\begin{tabular}{lccc}
 \multicolumn{1}{c}{} & \multicolumn{1}{c}{units}
&\multicolumn{1}{c}{$^4$He}          &\multicolumn{1}{c}{$^3$He }\\
 \hline
$\lambda_m$           &nm           &1265.615 9(4) &1265.683 9(2)\\
$\alpha_1(\lambda_m)$ &a.u.         &1151.590(4)  &1152.221(6)\\
$E_{r}$               &$\mu$K       &1.493      &1.982 \\
P                     &W            &0.905      &1.201\\
$\tau_{sc}$           &s            &4.596      &3.459\\
\end{tabular}
\end{ruledtabular}
\end{table}

The magic wavelength with a positive dynamic dipole polarizability can be used to design an ODT referring to Eq.~(\ref{e6}). From Tables~\ref{t3} and \ref{t4}, it is seen that among the magic wavelengths $\lambda_{m1}$, $\lambda_{m2}$, and $\lambda_{m3}$, only the dynamic polarizability at the magic wavelength around 1265 nm is positive. Including the finite nuclear mass and relativistic effects, the results of this particular magic wavelength are given to be 1265.615 9(4) nm for the $2\,^3S_1(M=\pm1)\rightarrow3\,^3S_1(M=0)$ transition of $^4$He, and 1265.683 9(2) nm for the $2\,^3S_1(F=3/2, M_F=\pm3/2)\rightarrow3\,^3S_1(F=3/2, M_F=\pm1/2)$ hyperfine transition of $^3$He. The corresponding dynamic dipole polarizabilities for $^4$He and $^3$He are 1151.590(4) and 1152.221(6) a.u., respectively. These large dynamic dipole polarizabilities indicate that the magic wavelength around 1265 nm might be useful for the design of an ODT with further analysis.

On the one hand, the design of an ODT needs enough depth to capture a certain number of atoms. Generally, a natural scale for the minimum depth of an optical trap as required for efficient loading is set to be a few of 10$T_{r}$, where $T_{r}= 2E_{r}$ is the recoil temperature, and $E_{r}=h^2/2m\lambda_m^2$ is the photon recoil energy~\cite{grimm00}. The larger photon recoil energy, the deeper trap depth and the higher laser power required. The photon recoil energy at the 1265 nm magic wavelength is calculated to be $E_{r}$=1.493 $\mu$K for $^4$He and 1.982 $\mu$K for $^3$He. Supposing the trap depth as low as to be 20$E_r$, to obtain the required laser power for creating this supposed trap depth conservatively, the focused laser beam with a large beam waist of 100 $\mu$m is needed. According to Eq.~(\ref{e6}), we obtain the required laser powers for different transitions, that are listed in the second to the last line of Table~\ref{t5}. The required laser powers are about 0.9 W for $^4$He and 1.2 W for $^3$He, which are feasible under current advanced laser technology. Therefore we believe that for the $2\,^3S_1\rightarrow3\,^3S_1$ transition, using the magic wavelength at 1265 nm with the laser power around 1 W can create a trap depth of about 30 $\mu$K for $^4$He and 40 $\mu$K for $^3$He. %($\sim$ 30 $\mu$K)

On the other hand, the design of an ODT needs small atomic scattering rate, since the low scattering rate ensure enough time for exciting the He atom from the $2\,^3S_1$ state to the $3\,^3S_1$ state. Using the calculated trapping beam power $P$ that can creat the supposed 20$E_{r}$ deep trap, we estimate the scattering rates for $^4$He and $^3$He isotopes at the magic wavelength around 1265 nm, then the trapping lifetimes are also obtained, which are given in the last line of Table~\ref{t5}. The trapping lifetimes are about 4.6 $s$ for $^4$He and 3.5 $s$ for $^3$He, which are long enough to perform spectral measurement of the forbidden $2\,^3S_1\rightarrow 3\,^3S_1$ transition.

\subsection{Suppressing ac Stark shift in two-photon transition}
%====================================================================

\begin{figure}[!htbp]
\includegraphics[width=0.49\textwidth]{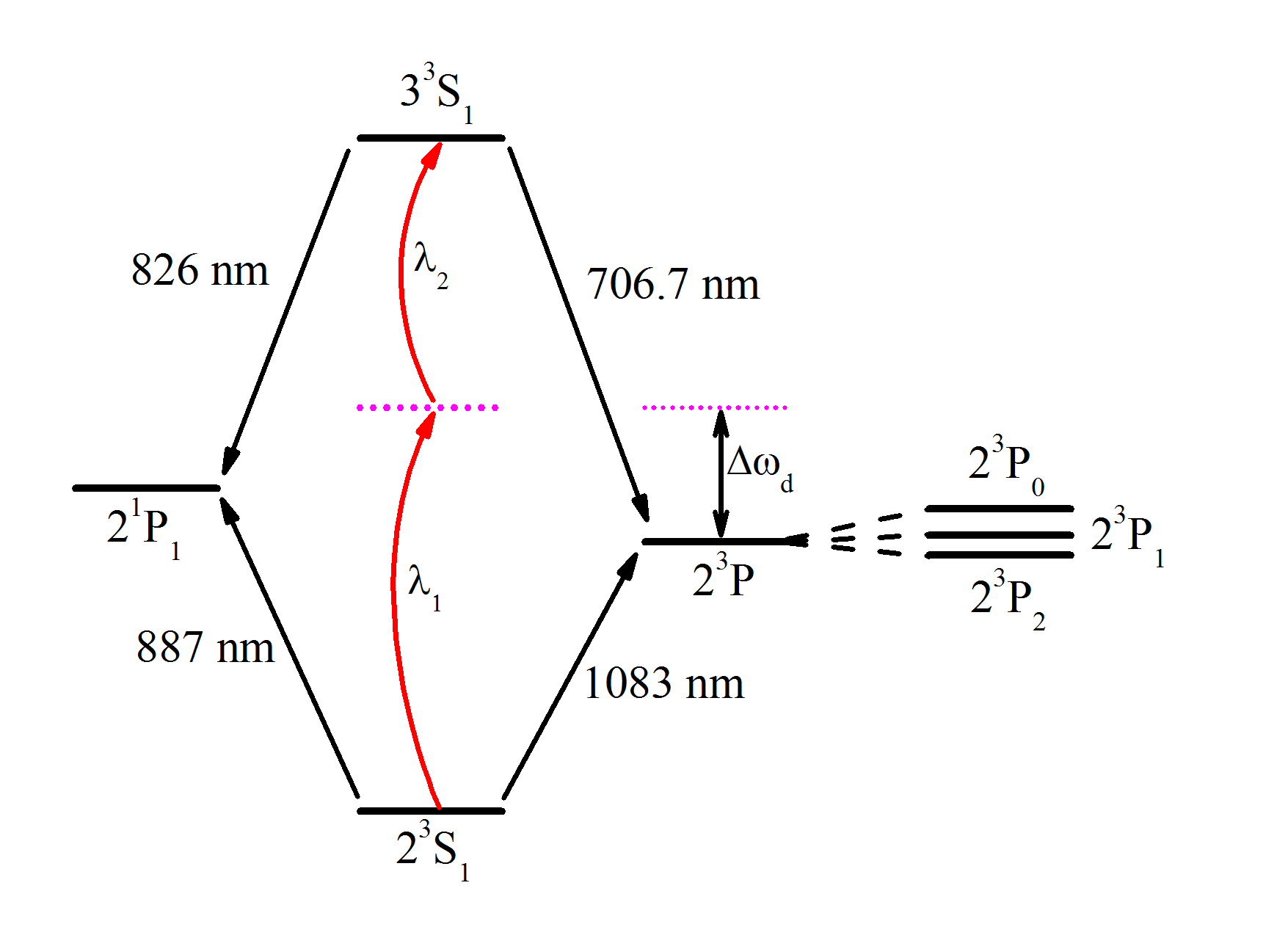}
\caption{\label{f2}(Color online) Energy level diagram indicating the $2\,^3S_1\rightarrow 3\,^3S_1$ two-photon transition of $^4$He excited by two lasers with wavelengths of $\lambda_1$ and $\lambda_2$. The detuning frequency of $\Delta \omega_{d}$ represents the relative position of the virtual state to the real $2\,^3P$ state. The single-photon transition wavelength of $2\,^3S_1\rightarrow 3\,^3S_1$ is 427.7 nm. Other relevant transition wavelengths are indicated. }
\end{figure}

It's seen from Sec. III C, once we use the laser with the magic wavelength around 1265 nm to build an ODT, the ac Stark shift in the $2\,^3S_1\rightarrow 3\,^3S_1$ transition caused by the trapping beam can be minimized to zero. However, the ac Stark shift caused by the probe laser has become the focus in the spectroscopy measurement of the $2\,^3S_1\rightarrow 3\,^3S_1$ transition. In this subsection, we will propose a two-photon excitation scheme to suppress the ac Stark shift in the probing process. The two-photon transition of $2\,^3S_1\rightarrow 3\,^3S_1$ in $^4$He is shown graphically in Fig.~\ref{f2}. This two-photon transition is excited by two different lasers with wavelengths of $\lambda_1$ and $\lambda_2$. The detuning frequency of $\Delta \omega_{d}$ indicates the relative position of the virtual state to the real $2\,^3P$ state.

\begin{figure}[!htbp]
\includegraphics[width=0.49\textwidth]{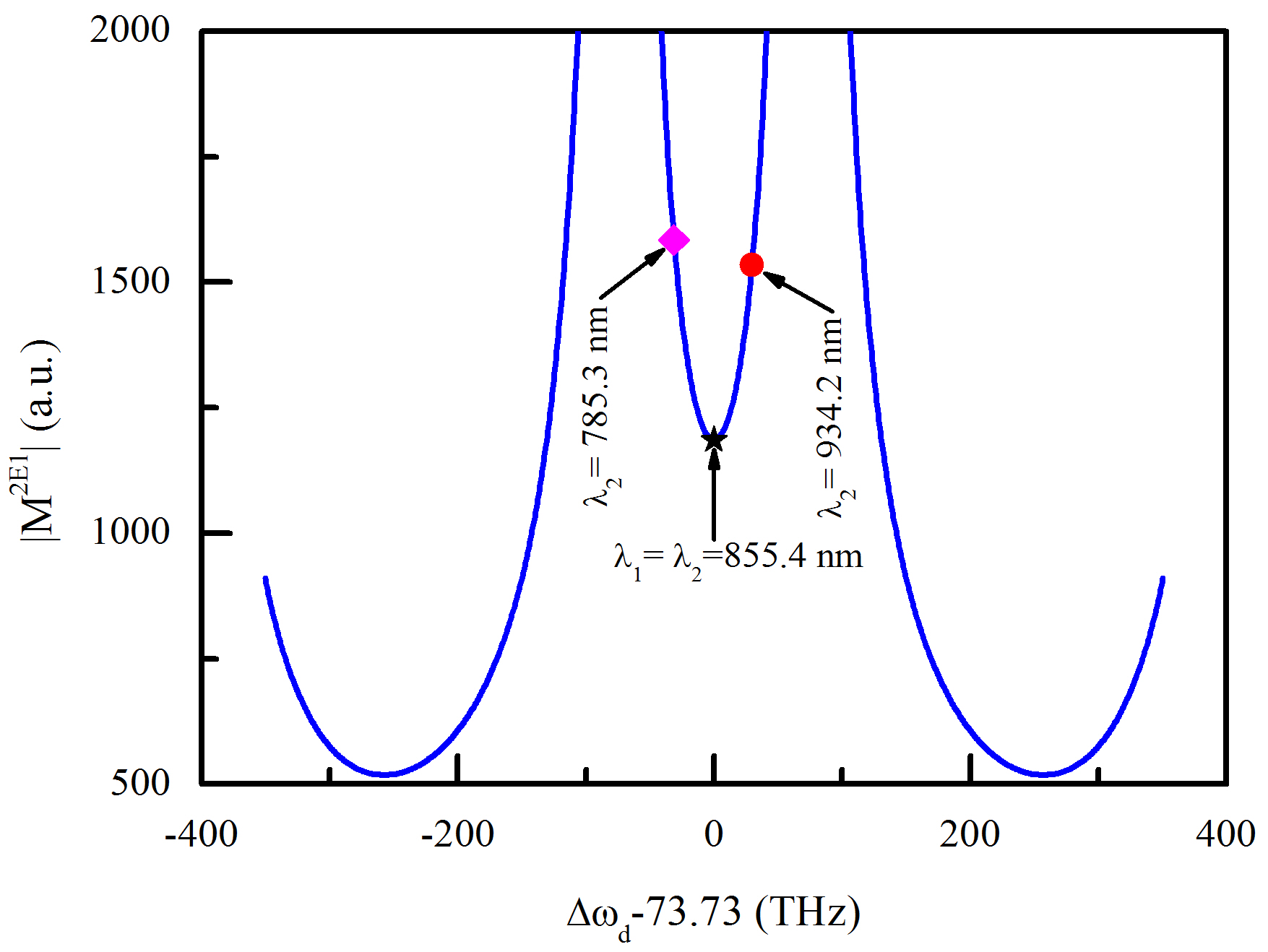}
\caption{\label{f4}(Color online) The transition amplitude $|M^{2E1}|$ of the $2\,^3S_1\rightarrow 3\,^3S_1$ two-photon transition of $^4$He as a function of the detuning frequency $\Delta \omega_d$. The two-photon transition is excited by two different lasers with wavelengths of $\lambda_1$ and $\lambda_2$. Three positions marked by arrows from left to right indicate the cases with $\Delta \omega_d=$ 42.46, 73.73, and 103.31 THz. For $\Delta \omega_d=$42.46 and 103.31 THz, $\lambda_2$ are respectively the 785.3 and 934.2 nm magic wavelengths, and for $\Delta \omega_d=$73.73 THz, two lasers have equal wavelengths. }
\end{figure}

\begin{table*}[!htbp]
\caption{\label{t6} The detuning frequency $\Delta \omega_d$, the two-photon transition amplitudes $|M^{2E1}|$, the laser wavelengths $\lambda_1$ and $\lambda_2$ exciting the two-photon transition, and the differences of dynamic dipole polarizabilities, $\Delta\alpha_1(\lambda_i)$, between the $3\,^3S_1(M=0)$ and $2\,^3S_1(M=\pm1)$ states of $^4$He at different laser wavelengths $\lambda_i\,,i=1,2$. }%For polarizabilities, the first one in brackets refers to the polarizability at $\lambda_1$, and the second one refers to the polarizability at $\lambda_2$.
\begin{ruledtabular}
\begin{tabular}{llllcc}
 \multicolumn{1}{c}{$\Delta \omega_d$(THz)}
&\multicolumn{1}{c}{$|M^{2E1}|$(a.u.)}  &\multicolumn{2}{c}{$\lambda_i$ (nm)}
&\multicolumn{2}{c}{$\Delta \alpha_1(\lambda_i)$(a.u.)}\\
 \cline{3-4}     \cline{5-6}
 \multicolumn{1}{c}{} &\multicolumn{1}{c}{}
&\multicolumn{2}{c}{($\lambda_1, \lambda_2$)}
&\multicolumn{1}{c}{$\Delta \alpha_1(\lambda_1)$} &\multicolumn{1}{c}{$\Delta \alpha_1(\lambda_2)$}\\
 \hline
103. 31 &1534.0  &\multicolumn{2}{c}{(788.8, 934.2)}     &$-$108.51               &0\\
73.73   &1185.7  &\multicolumn{2}{c}{(855.4, 855.4)}     &70.45                   &70.45\\
42.46   &1583.5  &\multicolumn{2}{c}{(939.2, 785.3)}     &$-$220.54               &0\\
\multicolumn{6}{c}{}\\
100  &1450.5  &\multicolumn{2}{c}{(795.8, 924.4)}     &$-$152.99                &108.24\\
10   &6347.5  &\multicolumn{2}{c}{(1045.5, 723.8)}    &3548.4                   &$-$1122.6\\
1    &63 804  &\multicolumn{2}{c}{(1079.4, 708.4)}    &4.15$\times10^{4}$       &$-$1.08$\times10^{4}$\\
0.1     &6.79$\times10^{5}$  &\multicolumn{2}{c}{(1082.9, 706.9)}  &4.25$\times10^{5}$  &$-$1.23$\times10^{5}$\\
0.01    &6.7$\times10^{6}$   &\multicolumn{2}{c}{(1083.3, 706.7)}  &4.8$\times10^{6}$   &$-$5.46$\times10^{5}$\\
\end{tabular}
\end{ruledtabular}
\end{table*}

To discuss the feasibility of the proposed scheme mentioned above, we first calculate the two-photon transition amplitudes at different detuning frequencies. Fig.~\ref{f4} plots the transition amplitudes $|M^{2E1}|$ of the $2\,^3S_1\rightarrow 3\,^3S_1$ two-photon transition in $^4$He at different detuning frequencies $\Delta \omega_d$, and several selected values of $|M^{2E1}|$ are listed in Table~\ref{t6}. As seen from Fig.~\ref{f4}, the curve of $|M^{2E1}|$ is symmetric with respect to $\Delta \omega_d=$ 73.73 THz, corresponding to conventional single-color case where the two-photon transition absorbs two equal-frequency photons.
From this symmetry position, with the decrease (or increase) of $\Delta \omega_d$, i.e., the virtual state is increasingly near to (or far away from) the real $2\,^3P$ state, the transition amplitude is enhanced significantly. For $\Delta \omega_d=42.46$ THz and $\Delta \omega_d=103.31$ THz marked by magenta diamond and red solid circle respectively in Fig.~\ref{f4}, both of the two-photon transition amplitudes are over $1.5\times10^3$ a.u., which are 192 times larger than the transition amplitude of 7.8 a.u. for the H($1s\rightarrow 2s$) two-photon transition at $\lambda_1=\lambda_2=$~243 nm~\cite{haas06}. In addition, the non-resonant two-photon decay rate is estimated to be 0.065 $s^{-1}$, that is at least six orders of magnitude larger than the magnetic dipole transition rate~\cite{thomas20}. These calculations provide a theoretical support for the feasibility of two-photon spectroscopy measurement of the He($2\,^3S_1\rightarrow 3\,^3S_1$) transition.

We next evaluate the ac Stark shift in the $2\,^3S_1\rightarrow 3\,^3S_1$ two-photon transition. While the two lasers with wavelengths $\lambda_1$ and $\lambda_2$ drive the two-photon transition, to the leading order in laser intensity and the fine-structure constant, the ac Stark shift can be calculated according to the following formula,
%\begin{widetext}
\begin{eqnarray}
\Delta E_{ac}=-\dfrac{1}{c_0\varepsilon_0}\left[\Delta \alpha_1(\lambda_1)I_1+\Delta \alpha_1(\lambda_2)I_2\right]\,,\label{e15}
\end{eqnarray}
%\end{widetext}
where the units are in SI, $\Delta \alpha_1(\lambda_i)$ is the difference of the electric dipole polarizabilities between the $3\,^3S_1(M=0)$ and $2\,^3S_1(M=\pm1)$ states, and the laser intensity $I_i=c\varepsilon_0F_i^2/2$, with $F_i$ being the electric field amplitude and $i=1,\,2$. Compared to the E1 contribution, the next-order dynamic electric quadrupole (E2) and magnetic dipole polarizabilities are respectively suppressed by a factor of $(\alpha \omega)^{2}$ and a factor of $\alpha^2$~\cite{porsev04a, porsev18}, and there are no resonant contributions
for the E2 and M1 polarizabilities at the frequencies of interest, so contributions from other multipole polarizabilities are neglected under the present calculations.

At different detuning frequencies, the laser wavelengths $\lambda_1$ and $\lambda_2$ exciting the $2\,^3S_1\rightarrow 3\,^3S_1$ two-photon transition, and the corresponding differences of dynamic polarizabilities between the $3\,^3S_1(M=0)$ and $2\,^3S_1(M=\pm1)$ states of $^4$He are given in Table~\ref{t6} as well. We can see that with $\Delta \omega_d$ decreasing from 100 THz to 0.01 THz, the difference of polarizabilities between the $3\,^3S_1$ and $2\,^3S_1$ states increases by three to four orders of magnitude, which will result in a large variation of ac Stark shift in the two-photon transition when $\Delta \omega_d$ varies. For $\Delta \omega_d=$73.73 THz, the total ac Stark shift will be twice of the Stark shift in the $\lambda_1$ laser field. For $\Delta \omega_d=$42.46 and 103.31 THz, the total ac Stark shift is only related to the $\lambda_1$ laser field, and the ac Stark shift induced by the $\lambda_2$ laser will vanish since $\lambda_2$ is a magic wavelength given in Sec. III B.

\begin{figure}[!htbp]
\includegraphics[width=0.49\textwidth]{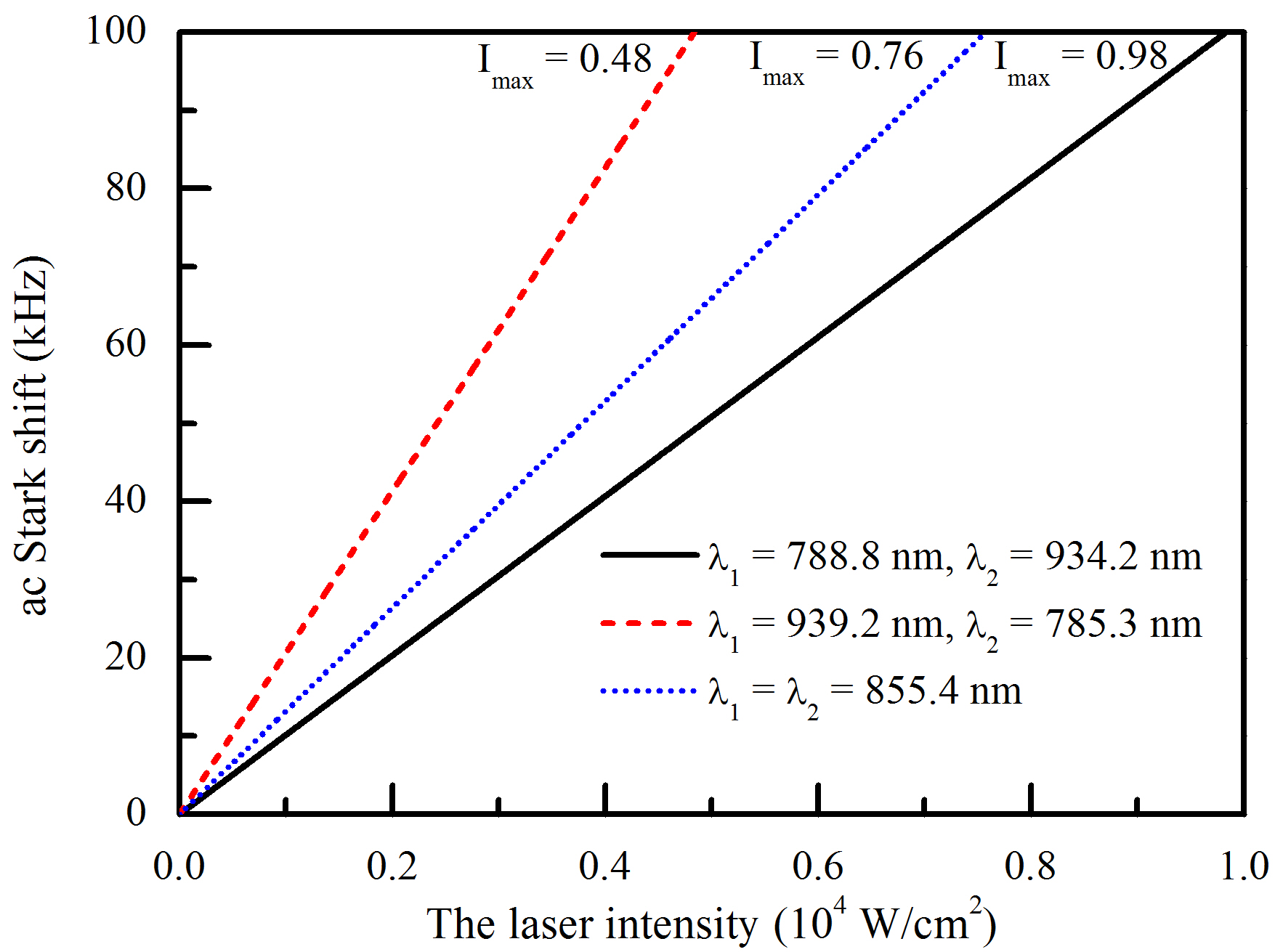}
\caption{\label{f3}(Color online) The ac Stark shifts in the $2\,^3S_1\rightarrow 3\,^3S_1$ two-photon transition of $^4$He at different intensities of the $\lambda_1$ laser. $I_{max}$ indicates the maximum intensity of the $\lambda_1$ laser when ac Stark shift equals 100 kHz. The two-photon transition is excited by two lasers with wavelengths of $\lambda_1$ and $\lambda_2$. The ac Stark shifts are estimated using dynamic dipole polarizabilities of $2\,^3S_1(M=\pm1)$ and $3\,^3S_1(M=0)$.}
\end{figure}

The ac Stark shifts in the $2\,^3S_1\rightarrow 3\,^3S_1$ two-photon transition of $^4$He as the $\lambda_1$ laser intensity changed are plotted in Fig.~\ref{f3}. The maximum laser intensity $I_{max}$ is given for suppressing the total ac Stark shift to be 100 kHz. We can see that to suppress the ac Stark shift at the same level, the laser intensity for $\Delta \omega_d=$ 103.31 THz (corresponding to $\lambda_1=788.8$ nm and $\lambda_2=934.2$ nm) is largest, which will make it easier to drive the two-photon transition. Moreover, for $\Delta \omega_d=$ 103.31 THz, as long as the $\lambda_1$ laser intensity does not exceed $1\times 10^4 ~W/cm^2$, the total ac Stark shift caused by probe beams will be less than 100 kHz, which would be about 70-fold smaller than the single-photon measurement~\cite{thomas20}. With this laser intensity, the collision heating caused by the probe beam is about 0.5 $\mu K s^{-1}$, which will not destroy the stability of the ODT. Also the laser intensity within the proposed maximum limit of $1\times 10^4 ~W/cm^2$ is high enough to drive the two-photon transition, since the laser intensity in the $2\,^3S_1\rightarrow 3\,^3S_1$ single-photon transition experiment is $3.86\times 10^3 ~W/cm^2$~\cite{thomas20} and we mentioned earlier that the two-photon transition rate is much larger than the single-photon M1 transition rate. Therefore, we suggest using these two lasers with wavelengths of $\lambda_1=788.8$ nm and $\lambda_2=934.2$ nm that are also well detuned from electric dipole transition frequencies, to realize the two-photon transition. The values for the $\lambda_2$ magic wavelength are 934.234 5(2) nm and 934.255 4(4) nm for $^4$He$(2\,^3S_1, M=\pm1\rightarrow3\,^3S_1, M=0)$ and $^3$He$(2\,^3S_1, F=3/2, M_F=\pm3/2\rightarrow3\,^3S_1, F=3/2, M_F=\pm1/2)$, respectively.
%, with QED correction of $-$0.001 05(2) nm included.

\begin{figure}[!htbp]
\includegraphics[width=0.49\textwidth]{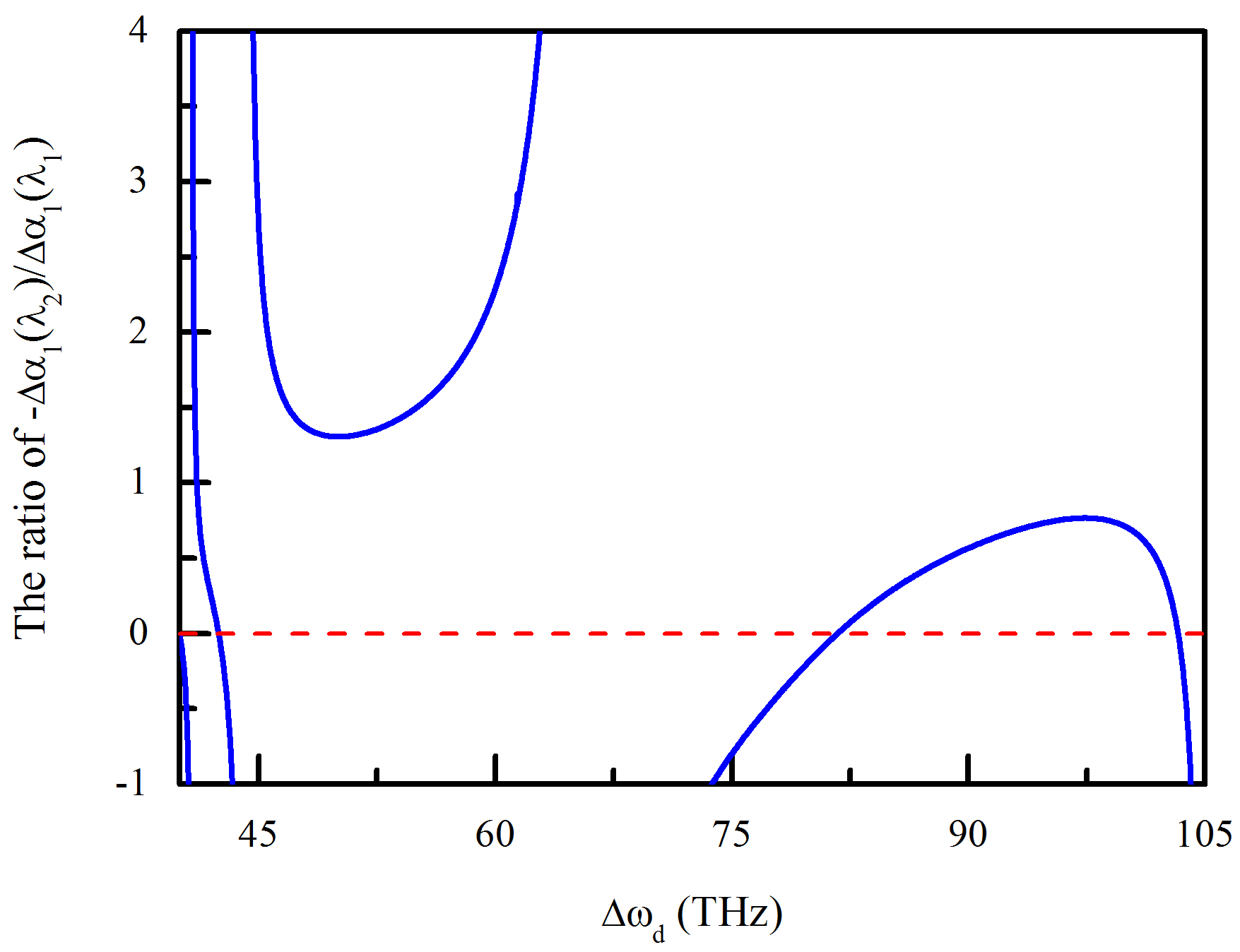}
\caption{\label{f5}(Color online) The ratios of $-\Delta \alpha_1(\lambda_2)/\Delta \alpha_1(\lambda_1)$ at different detuning frequencies $\Delta \omega_d$. $\Delta \alpha_1(\lambda_1)$ and $\Delta \alpha_1(\lambda_2)$ are the differences of polarizabilities between the $3\,^3S_1(M=0)$ and $2\,^3S_1(M=\pm1)$ states of $^4$He at the wavelengths of $\lambda_1$ and $\lambda_2$, respectively. The red dashed line is a horizonal zero line.}
\end{figure}

In addition, from Eq.~(\ref{e15}) we see that when $-\Delta \alpha_1(\lambda_2)/\Delta \alpha_1(\lambda_1)$ is a positive value, the ac Stark shift can be cancelled by adjusting the laser intensity ratio $I_1/I_2$ equal to $-\Delta \alpha_1(\lambda_2)/\Delta \alpha_1(\lambda_1)$. Gerginov and Beloy have demonstrated this method on the $5s\,^2S_{1/2}\rightarrow 5d\,^2D_{5/2}$
two-photon transition in $^{87}$Rb~\cite{gerginov18}. Considering comprehensively the two-photon transition amplitude and the residual first-order Doppler broadening due to the unequal laser wavelengths, for the $2\,^3S_1\rightarrow 3\,^3S_1$ two-photon transition of helium, the ratios of $-\Delta \alpha_1(\lambda_2)/\Delta \alpha_1(\lambda_1)$ only at $\Delta \omega_d=$ 40$\sim$105 THz are shown in Fig.~\ref{f5}. It is seen that for $\Delta \omega_d$ in the region of 82$\sim$103 THz, the ratios are positive and less than one, so the ac Stark shift cancellation method with an appropriate intensity ratio can also be used in the He($2\,^3S_1\rightarrow 3\,^3S_1$) two-photon transition. For example, driving the two-photon transition with the intensity ratio of $I_1/I_2=0.707$ will achieve zero ac Stark shift for $\Delta \omega_d=$ 100 THz given in Table~\ref{t6}; and a 0.1\% intensity ratio change leads to a 1.2-fold increase in the total ac Stark shift.

\section{conclusion}

%We have calculated the static and dynamic dipole polarizabilities for the $3\,^3S_1$ state of $^4$He and $^3$He isotopes. The static scalar dipole polarizabilities for the $3\,^3S_1$ state of $^4$He and $^3$He are obtained to be 7940.361(4) a.u. and 7942.121(6) a.u. in the present RCI calculations; the $\alpha^3$-order QED correction of 0.183(2) a.u. is acquired by using the perturbation method.
We have determined magic wavelengths for the $2\,^3S_1\rightarrow 3\,^3S_1$ forbidden transition in $^4$He and $^3$He isotopes, and proposed a new experimental scheme for suppressing the ac Stark shift in the $2\,^3S_1\rightarrow 3\,^3S_1$ transition frequency measurement. For $^4$He, the 1265.615 9(4) nm magic wavelength can be used to design an optical dipole trap, which can create a 20$E_{r}$ trap depth with the laser power of 0.9 W and has 4.6 $s$ trapping lifetime. Furthermore, the 934.234 5(2) nm magic wavelength is suggested as the $\lambda_2$ laser to excite the two-photon process for the $2\,^3S_1\rightarrow 3\,^3S_1$ transition, and the ac Stark shift would be reduced to about 70-fold smaller compared to the single-photon transition, as long as the intensity of the $\lambda_1$ laser does not exceed $1\times10^4~W/cm^2$. Similarly, for $^3$He, the 1265.683 9(2) nm magic wavelength can be used to design an ODT and the 934.255 4(4) nm magic wavelength can be as the $\lambda_2$ laser to realize the $2\,^3S_1\rightarrow 3\,^3S_1$ two-photon transition process.
Alternatively, for detuning frequencies relative to $2\,^3P$ state in the region of 82$\sim$103 THz, driving the two-photon transition with appropriate intensity ratios will achieve zero ac Stark shift.
We expect that our proposal can improve the measured precision of He($2\,^3S_1\rightarrow3\,^3S_1$) transition frequency.

\section{acknowledgement}

We would like to thank Cheng-Bin Li, Lin-Qiang Hua, and Yu Robert Sun for their helpful comments on the manuscript, and thank Prof. G. W. F. Drake for instructive discussions regarding the two-photon transitions.
This work was supported by the National Natural Science Foundation of China under Grants No. 11704398 and No. 11774386, by the Strategic Priority Research Program of the Chinese Academy of Sciences, Grants No. XDB21010400 and No. XDB21030300, by the National Key Research and Development Program of China under Grant No. 2017YFA0304402, and by the Hubei Province Science Fund for Distinguished Young Scholars No. 2019CFA058.
%for discussions about trapping atoms in optical traps

%\bibliography{positron}

\end{document}